\documentclass{article}

\usepackage[english]{babel}

\usepackage[letterpaper,top=2cm,bottom=2cm,left=3cm,right=3cm,marginparwidth=1.75cm]{geometry}

\usepackage{amsmath}
\usepackage{graphicx}
\usepackage[colorlinks=true, allcolors=blue]{hyperref}
\usepackage{float}
\usepackage{multirow} 
\usepackage{graphicx}
\usepackage{array}  
\usepackage{csquotes}
\usepackage[backend=biber,style=numeric-comp,sorting=none,maxnames=6,minnames=6]{biblatex}
\usepackage[bottom]{footmisc}
\usepackage{authblk} 
\usepackage{subcaption}

\title{A recent evaluation on the performance of LLMs on radiation oncology physics using questions of randomly shuffled options}

\author[1]{Peilong Wang, PhD}
\author[1]{Jason Holmes, PhD}
\author[2]{Zhengliang Liu, MS}
\author[3]{Dequan Chen, PhD}
\author[2]{Tianming Liu, PhD}
\author[1]{Jiajian Shen, PhD}
\author[1]{Wei Liu, PhD\thanks{Corresponding author}}

\affil[1]{Department of Radiation Oncology, Mayo Clinic, Phoenix, AZ 85054}
\affil[2]{School of Computing, University of Georgia, Athens, GA 30602}
\affil[3]{Department of Radiology, Mayo Clinic, Rochester, MN 55905}
\date{}

\begin{document}
\maketitle

\begin{abstract}
\noindent Purpose:
We present an updated study evaluating the performance of large language models (LLMs) in answering radiation oncology physics questions, focusing on the recently released models.

\vspace{1em} 

\noindent Methods:
A set of 100 multiple-choice radiation oncology physics questions, previously created by a well-experienced physicist, was used for this study. The answer options of the questions were randomly shuffled to create "new" exam sets. Five LLMs -- OpenAI o1-preview, GPT-4o, LLaMA 3.1 (405B), Gemini 1.5 Pro, and Claude 3.5 Sonnet -- with the versions released before September 30, 2024, were queried using these new exam sets. To evaluate their deductive reasoning ability, the correct answer options in the questions were replaced with "None of the above." Then, the explain-first and step-by-step instruction prompts were used to test if this strategy improved their reasoning ability. The performance of the LLMs was compared with the answers from medical physicists.

\vspace{1em} 

\noindent Results:
All models demonstrated expert-level performance on these questions, with o1-preview even surpassing medical physicists with a majority vote. When replacing the correct answer options with 'None of the above', all models exhibited a considerable decline in performance, suggesting room for improvement. The explain-first and step-by-step instruction prompts helped enhance the reasoning ability of the LLaMA 3.1 (405B), Gemini 1.5 Pro, and Claude 3.5 Sonnet models.

\vspace{1em} 

\noindent Conclusion:
These recently released LLMs demonstrated expert-level performance in answering radiation oncology physics questions. 

\end{abstract}

\section{Introduction}
Large language models (LLMs) have advanced rapidly. On the one hand, the size of the data used for pre-training and the number of model parameters have grown a lot. For example, GPT-2 had 1.5 billion parameters~\cite{gpt2_webpage}, GPT-3 scaled up to 175 billion~\cite{NEURIPS2020_1457c0d6}, and GPT-4 is estimated to have even more~\cite{openai2024gpt4technicalreport}. On the other hand, the fine-tuning methods and prompt engineering strategies have advanced substantially~\cite{NEURIPS2022_b1efde53, hu2021loralowrankadaptationlarge}. Furthermore, agents and Retrieval-Augmented Generation (RAG) systems built on LLMs have seen considerable progress~\cite{scichinaiformatic, NEURIPS2020_6b493230}. Notable recent developments as of September 2024 include OpenAI o1-preview~\cite{o1_webpage}, GPT-4o~\cite{openai2024gpt4ocard}, LLaMA 3.1 (405B parameters)~\cite{grattafiori2024llama3herdmodels}, Gemini 1.5 Pro~\cite{geminiteam2024gemini15unlockingmultimodal}, and Claude 3.5 Sonnet~\cite{claude_webpage}, demonstrating state-of-art performance in overall language processing, reasoning, and diverse downstream applications. 

The rapid evolution of LLMs also renders prior performance evaluations outdated. As some LLMs cease providing services, new models are introduced, and existing versions are updated, studies published before may no longer accurately reflect the current state of LLM capabilities. A fresh evaluation is needed to address the dynamic landscape of LLM advancements.

In healthcare, LLMs have been explored for numerous potential applications~\cite{Liu2024a,LIU2023100045,zhangkai2024,HOLMES2024e515,li2024}. For their direct use in radiation oncology, unique challenges related to evaluation and validation arise due to the complexity and precision of treatment, which involves both clinical factors and physics considerations. Therefore, assessing the performance of LLMs in addressing questions related to radiation oncology physics is crucial. Such evaluations not only tell us how efficiently they process and reason about radiation oncology physics but also help us understand their limitations. In the past, several LLMs were evaluated on the 2021 American College of Radiology (ACR) Radiation Oncology In-Training Examination (TXIT), revealing that GPT-4-turbo achieved the highest score of 68.0\%, outperforming some resident physicians~\cite{doi:10.1089/aipo.2023.0007}. GPT-3.5 and GPT-4 were also assessed on Japan's medical physicist board examinations from 2018 to 2022, where GPT-4 demonstrated superior performance with an average accuracy of 72.7\%~\cite{Kadoya2024-pi}. To offer insights into the recently released state-of-art LLMs and build on our prior work~\cite{10.3389/fonc.2023.1219326}, we present here an updated study with refined methods evaluating their performance in radiation oncology physics.


We utilized the 100-question radiation oncology physics exam we developed based on the American Board of Radiology exam style~\cite{liu2024radiationoncologynlpdatabase}, and randomly shuffled the answer options to create "new" exam sets. We then queried the LLMs with these new exam sets and checked their ability to answer questions accurately. We also evaluated their deductive reasoning ability and tested whether the explain-first and step-by-step instruction prompts would improve their performance in reasoning tasks.

\section{Methods}
The 100-question multiple-choice examination on radiation oncology physics was created by our experienced medical physicist, following the official study guide of the American Board of Radiology. That exam includes 12 questions on basic physics, 10 questions on radiation measurements, 20 questions on treatment planning, 17 questions on imaging modalities and applications in radiotherapy, 13 questions on brachytherapy, 16 questions on advanced treatment planning and special procedures, and 12 questions on safety, quality assurance (QA), and radiation protection. 17 out of the 100 questions are math-based and require numerical calculation.

All the evaluated LLMs were queried with the exam questions through Application Programming Interface (API) services provided by their respective hosts, except LLaMA 3.1 (405B), an open-source LLM, which was hosted by us locally at our institution. All the LLMs used were the recently released version before September 30, 2024. The temperature was set to 0.1 for all LLMs to minimize variability in their responses\footnote{The temperature was set to 0.1 rather than 0 due to the different meanings of a temperature of 0 across different LLMs. To avoid potential unexpected behaviors from models, we set the lower bound to 0.1 rather than 0. }, with the exception of the OpenAI o1-preview, whose temperature was fixed at 1 and could not be changed by the user.

\subsection{Randomly shuffling the answer options}
\label{sec:random_shuffling}
Since it was difficult to know whether any LLM had been pre-trained using our previously published 100-question multiple-choice exam, we wrote Python code to randomly shuffle the answer options for the 100 multiple-choice questions five times. For each shuffle, we obtained a "new" 100-question multiple-choice exam set. We then queried all the LLMs five times (Trial 1 - Trial 5), each with a new exam set. Each question of the new exam set was queried individually. We checked the distribution of the correct answers' locations for the five new exams where the options were shuffled and confirmed that the distribution of the correct options is fairly random among A, B, C, D, or E (only 2 questions offered option E), as shown in the supplementary material.

The prompt we used for all the queries was as follows:
\begin{quote} "Please solve this radiation oncology physics problem:

[radiation oncology physics problem].
" \end{quote}
This allowed the LLMs to reason and answer freely. Table \ref{tab:prompt_plus_question} illustrates an example of the trials and how we queried the LLMs with the questions. For the responses generated by the LLMs, we utilized the LLaMA 3.1 (405B) model hosted locally to further extract the chosen answer options (letters A, B, C, D, or E) from free-form responses, thereby reducing some of the manual effort required to read and record them individually. We then conducted manual verification of the extracted options and obtained the final answer option sheet for all LLMs to compare with the ground truth answers. The accuracy of each LLM was reported as the mean score across the five trials, and the measurement uncertainty was reported as the standard deviation of the five trials, as shown in the following equations:
\[
\text{Mean} (\bar{x}) = \frac{1}{N} \sum_{i=1}^N x_i,
\]

\[
\sigma = \sqrt{\frac{1}{N-1} \sum_{i=1}^N (x_i - \bar{x})^2},
\]
where \( x_i \) represents each measurement.

The results of the LLMs' test scores were compared with the majority vote results from a group of medical physicists conducted in our previous study. The medical physicist group consisted of four experienced board-certified medical physicists, three medical physics residents, and two medical physics research fellows. For each question, the most common answer choice was selected as the group's answer. In case of a tie, one of the most common answer choices was chosen randomly.

\begin{table}[h]
\caption{Illustration of the prompts and questions of randomly shuffled options to evaluate LLMs' performance on answering radiation oncology physics questions.}
\label{tab:prompt_plus_question}
\resizebox{\textwidth}{!}{%
\begin{tabular}{|l|l|l|}
\hline
\textbf{Trial}   & \textbf{Prompt}                              & \textbf{Question}                                                                                                                                                                          \\ \hline
Trial 1 & \multirow{5}{*}{\vspace{-7cm} Please solve this radiation oncology physics problem:} & \begin{tabular}[c]{@{}l@{}}Which of the following particles cannot be accelerated by an electric field?\\ (a) Neutrons\\ (b) Protons\\ (c) Electrons\\ (d) Positrons\end{tabular} \\ \cline{1-1} \cline{3-3} 
Trial 2 &                                                                        & \begin{tabular}[c]{@{}l@{}}Which of the following particles cannot be accelerated by an electric field?\\ (a) Protons\\ (b) Neutrons\\ (c) Electrons\\ (d) Positrons\end{tabular} \\ \cline{1-1} \cline{3-3} 
Trial 3 &                                                                        & \begin{tabular}[c]{@{}l@{}}Which of the following particles cannot be accelerated by an electric field?\\ (a) Positrons\\ (b) Protons\\ (c) Electrons\\ (d) Neutrons\end{tabular} \\ \cline{1-1} \cline{3-3} 
Trial 4 &                                                                        & \begin{tabular}[c]{@{}l@{}}Which of the following particles cannot be accelerated by an electric field?\\ (a) Electrons\\ (b) Neutrons\\ (c) Protons\\ (d) Positrons\end{tabular} \\ \cline{1-1} \cline{3-3} 
Trial 5 &                                                                        & \begin{tabular}[c]{@{}l@{}}Which of the following particles cannot be accelerated by an electric field?\\ (a) Electrons\\ (b) Positrons\\ (c) Neutrons\\ (d) Protons\end{tabular} \\ \hline
\end{tabular}%
}
\end{table}

\subsection{Evaluating deductive reasoning ability}
Deductive reasoning ability refers to the cognitive process of logically analyzing information to draw specific conclusions from general premises. The multiple-choice question with the answer option "None of the above" can effectively evaluate the test-taker’s deductive reasoning ability, as it involves evaluating each option based on the information provided and ruling out incorrect choices contradicting known facts or logical outcomes to reach the correct answer. We therefore replaced the correct option in the exam with "None of the above." Since transformer-based LLMs predict the next word based on prior contexts, changing the correct option to "None of the above" removes a straightforward cue that might guide the model toward a known or patterned solution, thus forcing the LLMs to rely more on reasoning about the specific question and its options, rather than using surface-level lexical or statistical patterns that it may have learned.

\subsubsection{Replacing the correct option with "None of the above"}
\label{sec:replace_with_noa}
We developed Python code to replace the correct option with "None of the above" for all questions in the five new exam sets derived by random shuffling. For each trial (Trial 1 - Trial 5), we queried the LLMs with a set of exams in which both the correct option was "None of the above," and its location was randomly shuffled. We used the same prompt as in Sec. \ref{sec:random_shuffling} across all queries, and each question in an exam set was queried individually. This setup challenges the LLMs to avoid pattern-based answering and not rely on any single choice, but to process each answer option by reasoning. 

As before, we utilized the previously described processes for answer-option extraction and manual verification, as outlined in Sec. \ref{sec:random_shuffling}. The performance accuracy and uncertainty of each LLM were reported as the average score and standard deviation across all five trials. Due to this setup, these exams were not used to test humans, as this pattern can be easily recognized by human test-takers\footnote{Since each question was queried through the API individually, it is assumed that LLMs would not notice this pattern.}.

\subsubsection{Explain-first and step-by-step instruction}
To further check if explicitly asking the LLMs to explain first and then develop answers step-by-step (chain-of-thought) would improve their deductive reasoning ability~\cite{wei2023chainofthoughtpromptingelicitsreasoning}, we engineered the following prompt and queried the LLMs again with it:
\begin{quote}
"Please solve this radiation oncology physics problem:

[radiation oncology physics problem]

Please first explain your reasoning, then solve the problem step by step, and lastly provide the correct answer (letter choice)."
\end{quote}
We used the five exam sets and conducted the querying process both as described in Sec. \ref{sec:replace_with_noa}. All five LLMs were evaluated using this prompting strategy. Accuracy and uncertainty were reported. The results from this strategy were compared with the test results from original prompts, where no explanation or step-by-step answering was required, as described in Sec. \ref{sec:replace_with_noa}.

\section{Results}

\subsection{Results of exam sets with randomly shuffled options}
\label{sec:results_random_shuffling}
\begin{figure}[ht]
\centering
\includegraphics[width=\linewidth]{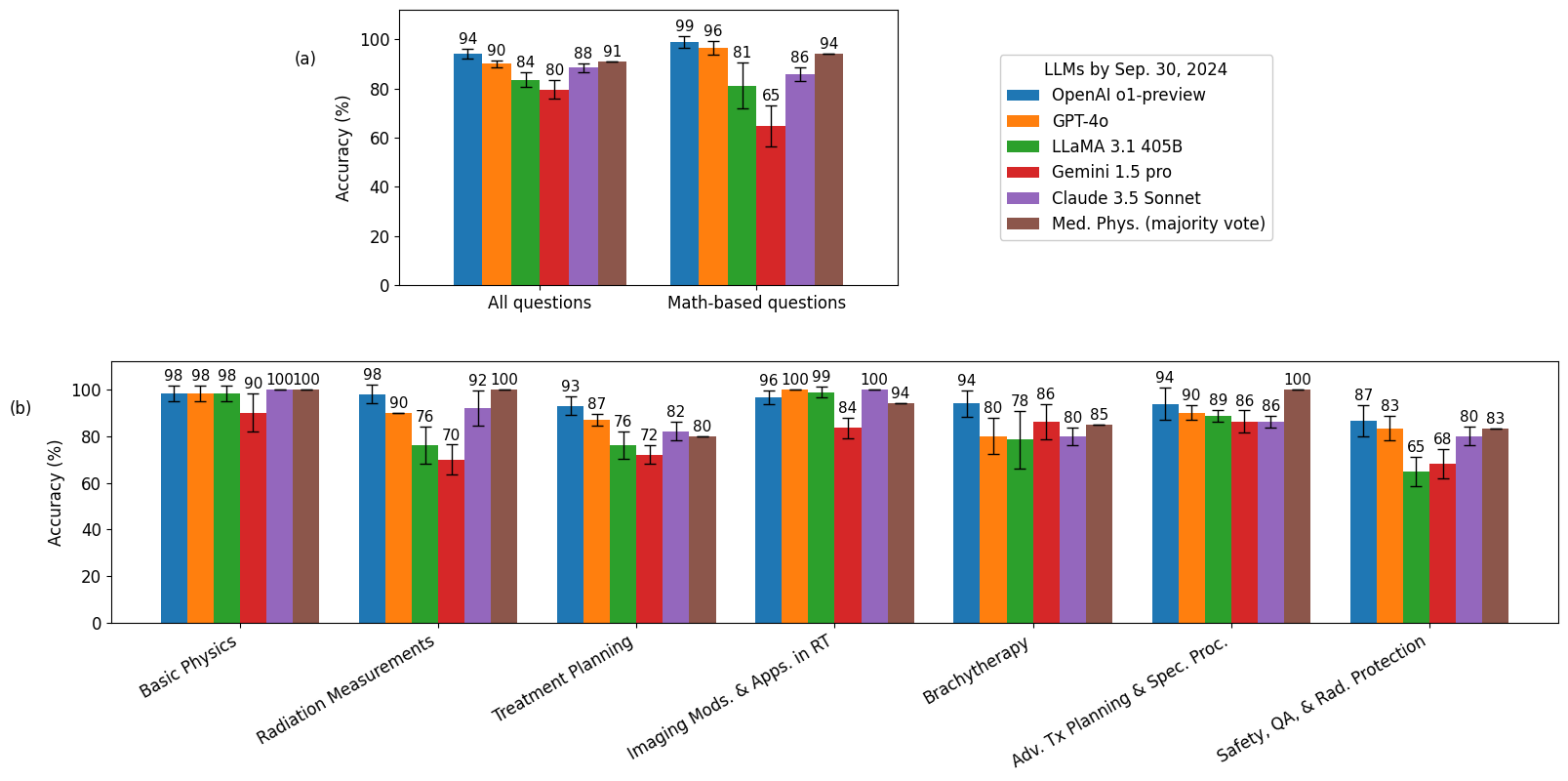}
\caption{\label{fig:evaluation_accuracy} Evaluation results using five exam sets where the answer options of the questions were randomly shuffled. Figure (a) illustrates the evaluation results for all questions and math-based questions, while Figure (b) presents the evaluation results broken down by different topics.}
\end{figure}

The evaluation results of the exam sets with options randomly shuffled are presented in Fig. \ref{fig:evaluation_accuracy}, where the height of each bar represents the mean test score, and the error bars indicate the standard deviations across five trials. All five LLMs exhibited strong performance, achieving mean test scores above 80\%, which suggests their performance on these exams is comparable to that of human experts. When compared to the majority vote results from the medical physics group, the OpenAI o1-preview model outperformed the medical physicists with a majority vote. For math-based questions, both the o1-preview and GPT-4o models surpassed the medical physicists with a majority vote.

\begin{figure}[H]
\centering
\includegraphics[width=\linewidth]{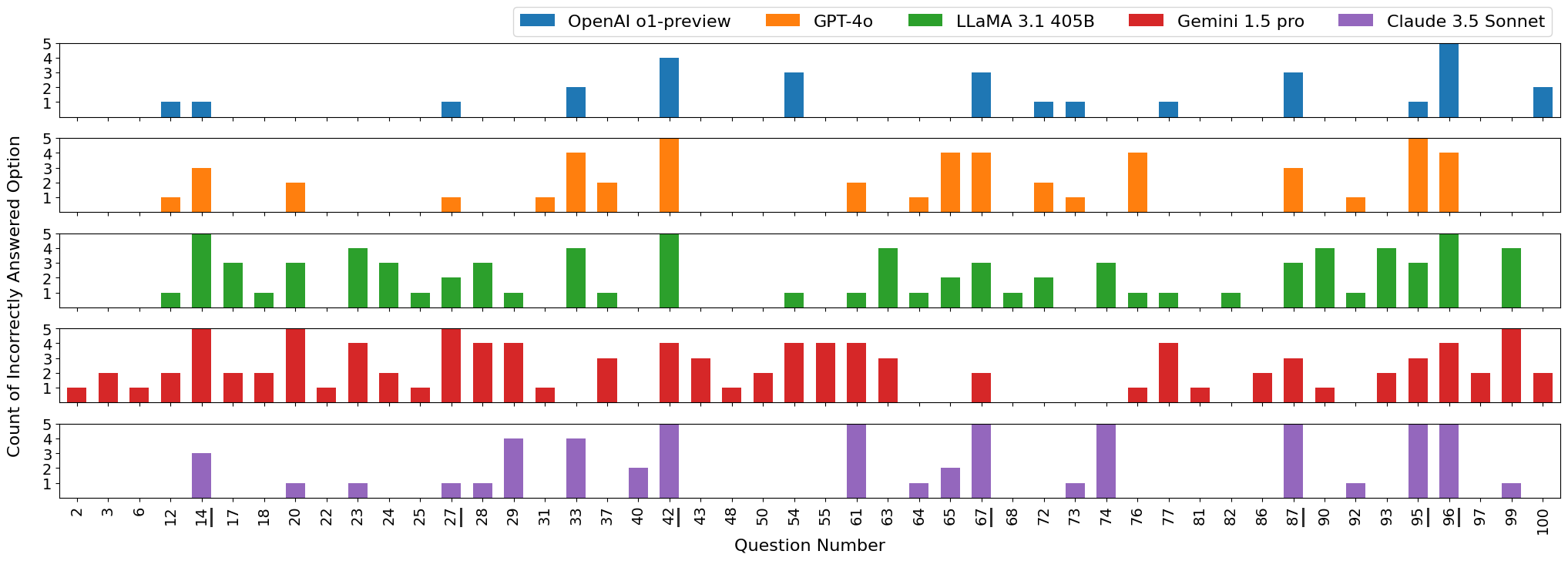}
\caption{\label{fig:dist_correct_wrong_options} Distribution of incorrectly selected answer options by each LLM across five exam sets. All answer options in the five exam sets were randomly shuffled. Questions that were correctly answered by all LLMs in all five trials are not shown in this figure. Questions 14, 27, 42, 67, 87, 95, and 96, which were commonly answered incorrectly by all LLMs at least once, are underlined in the figure.}
\end{figure}

The raw counts of incorrect responses by the LLMs are shown in Fig. \ref{fig:dist_correct_wrong_options}, where each color represents the incorrect answers by an LLM across trials. As observed, each LLM exhibited variability in answering questions across trials. Notably, the models also showed similarities in incorrectly answering certain questions. We analyzed the questions that were commonly answered incorrectly by all LLMs at least once across all five trials -- question numbers: 14, 27, 42, 67, 87, 95, and 96. Interestingly, only one of these questions was math-based, while the remaining seven were closely related to clinical medical physics knowledge, such as American Association of Physicists in Medicine (AAPM) Task Group (TG) reports and clinical experience. This observation suggests that current LLMs may still struggle with answering clinically focused radiation oncology physics questions. For example, question number 42 does not involve any calculations but instead focuses primarily on clinical hands-on experience.

\subsection{Results of LLMs' deductive reasoning ability}
\begin{figure}[ht]
\centering
\includegraphics[width=\linewidth]{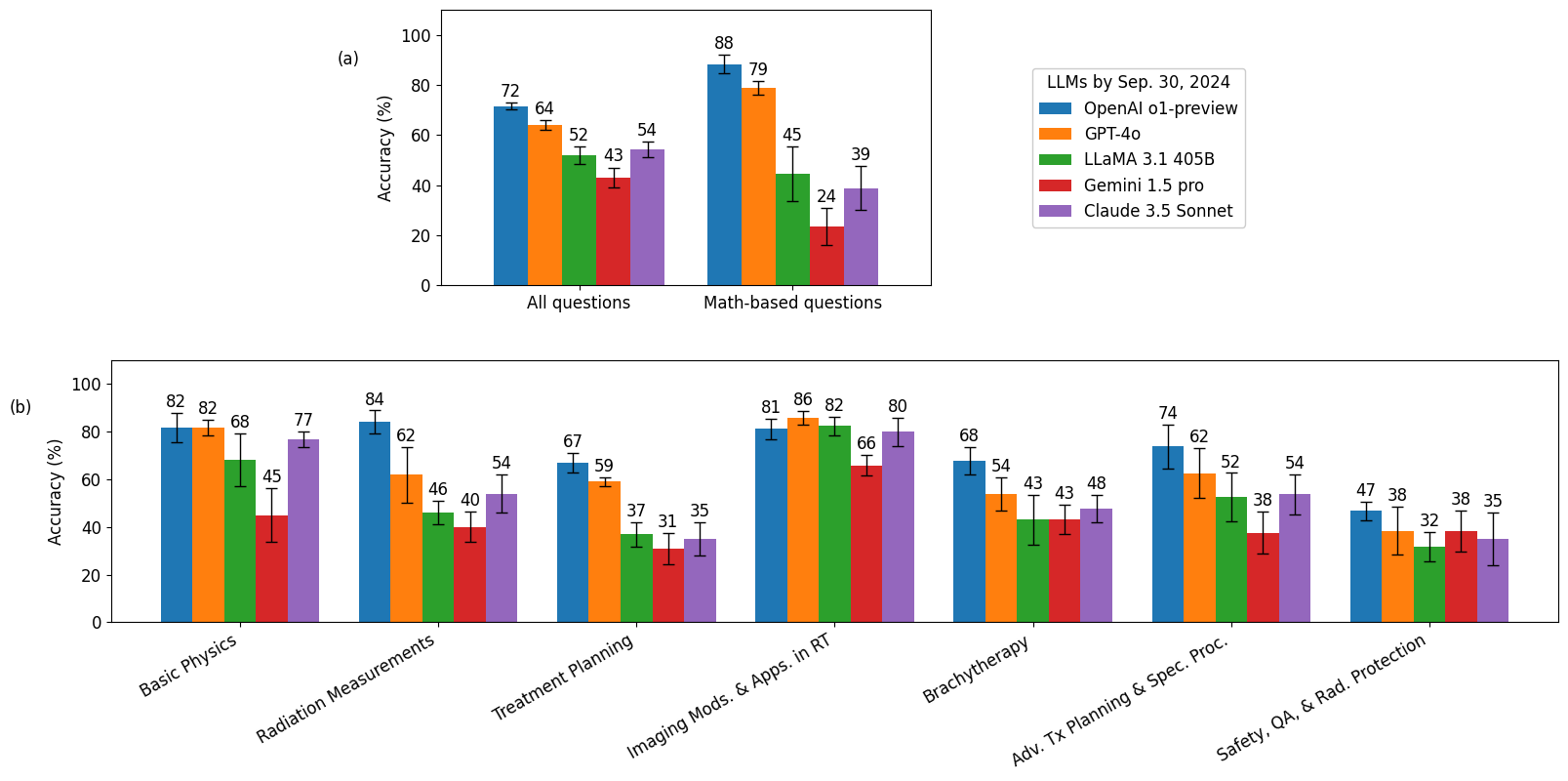}
\caption{\label{fig:evaluation_accuracy_noa} Deductive reasoning ability evaluation results for every LLM. The correct answer options were replaced with "None of the above" in all questions. Figure (a) illustrates the evaluation results for all questions and math-based questions, while Figure (b) presents the evaluation results broken down by different topics.}
\end{figure}

Fig. \ref{fig:evaluation_accuracy_noa} shows the results of the deductive reasoning ability tests, where the correct answer options were replaced with "None of the above" in all questions. Overall, all LLMs performed much more poorly compared to the results in Sec. \ref{sec:results_random_shuffling}. Given that transformer-based LLMs~\cite{NIPS2017_3f5ee243} were designed to predict the next word in a sequence, replacing the correct answers with "None of the above" would likely disrupt their pattern recognition abilities, thereby reducing their overall scores performed on the exam sets. Nonetheless, the OpenAI o1-preview and GPT-4o still outperformed the others, especially on math-based questions, indicating the strong reasoning ability of these two models.

\begin{figure}[H]
\centering
\includegraphics[width=\linewidth]{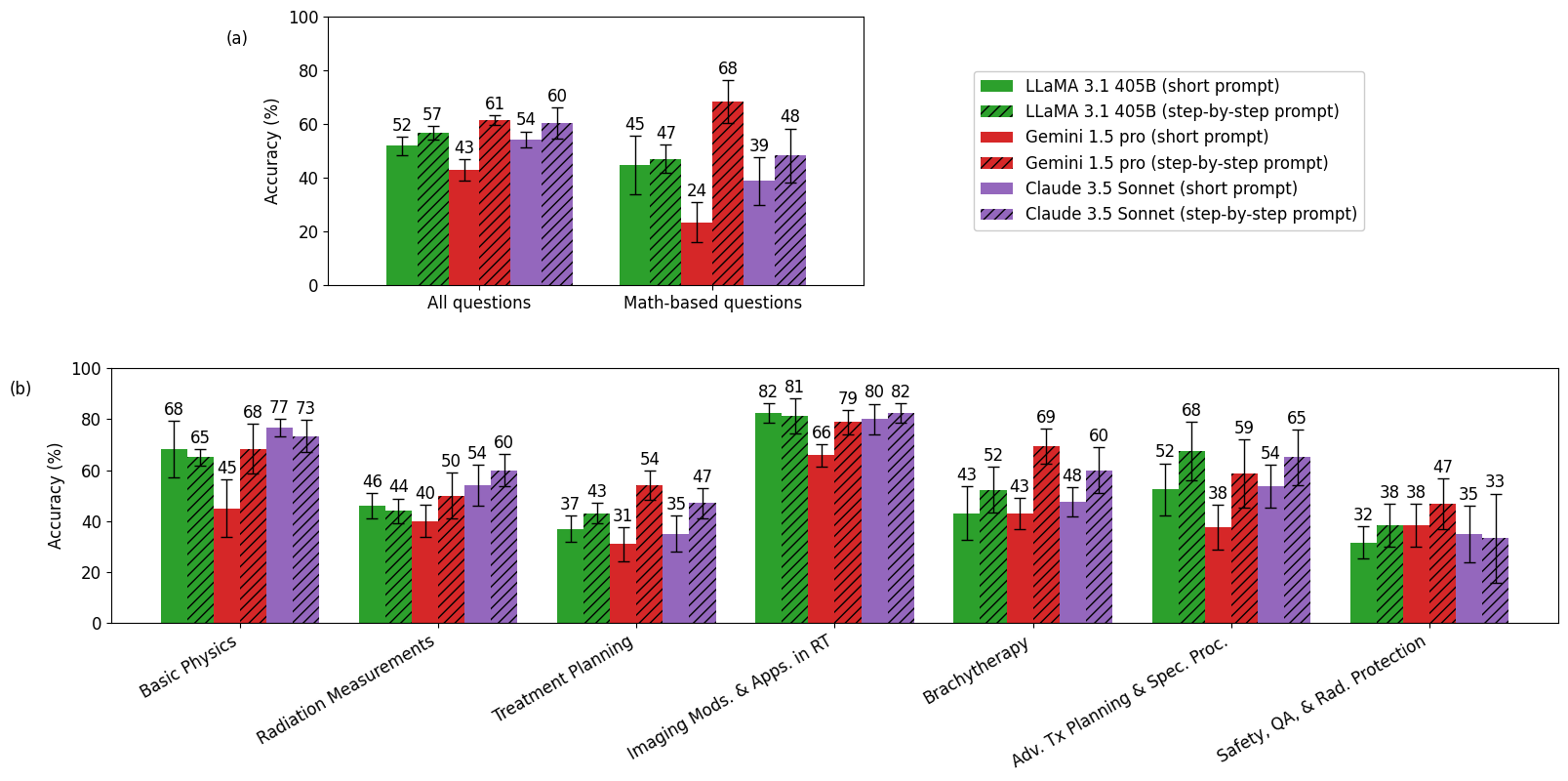}
\caption{\label{fig:evaluation_accuracy_noa_cot} Comparison of accuracy between the short prompt and the explain-first, step-by-step instruction prompt (chain-of-thought) for LLaMA 3.1 (405B), Gemini 1.5 Pro, and Claude 3.5 Sonnet. Figure (a) shows the comparison between the short prompt and the explain-first, step-by-step instruction prompt for all questions and math-based questions, while Figure (b) breaks down the comparison by topic. (Note: o1-preview and GPT-4o showed only about a 1\% overall difference, which was too small to be represented in this figure.)
}
\end{figure}

Fig. \ref{fig:evaluation_accuracy_noa_cot} compares the performance of LLaMA 3.1 (405B), Gemini 1.5 Pro, and Claude 3.5 Sonnet models with the original simple prompts and with the explain-first, step-by-step instruction prompts. Overall, all three models demonstrated improved reasoning ability with the latter prompting strategy. Notably, Gemini 1.5 Pro showed significant gains on math-based questions, increasing its score from 24\% to 68\%. The o1-preview and GPT-4o showed only about a 1\% overall difference, which was too small to be represented in this figure.\\

\section{Discussion}

\subsection{Improvement of performance on answering radiation oncology physics questions of the state-of-art LLMs over the past two years}
\begin{figure}[ht]
\hspace{2.3cm} 
\includegraphics[width=0.818\linewidth]{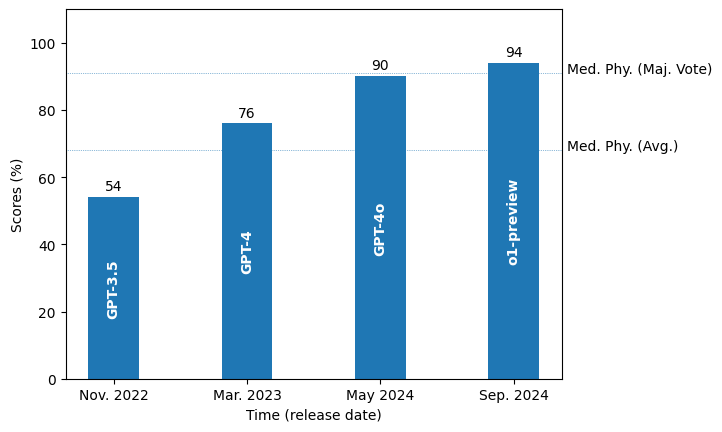}
\caption{\label{fig:growth figure} Growth of the state-of-art LLMs' performance in radiation oncology physics over the past two years. Two dotted lines mark the average score and the majority vote score of the medical physicists.}
\end{figure}

Over the past two years, our studies have observed a notable improvement in the performance of state-of-the-art LLMs on this highly specialized task -- answering radiation oncology physics questions, as shown in Fig. \ref{fig:growth figure}. Early versions of ChatGPT, like GPT-3.5 in late 2022~\cite{chatgpt_webpage}, scored around 54\%, showing clear gaps in domain-specific knowledge. With the introduction of GPT-4 in early 2023
, performance jumped to around 76\%, reflecting improvements in accuracy and understanding. Subsequent releases of the GPT-4o model and more recently the o1-preview (both in 2024), pushed scores even higher to 90\% and 94\% respectively, indicating increasing capabilities in radiation oncology physics. This steady improvement can be attributed to more extensive domain pre-training, increase of number of parameters, refined architectural updates, and enhanced fine-tuning techniques~\cite{gururangan-etal-2020-dont, kaplan2020scalinglawsneurallanguage}, all of which have led to improved understanding, stronger reasoning skills, and better alignment with expert-level knowledge. The evolution of these models over the last two years underscores the rapid growth of LLMs, suggesting their potential as useful tools in areas such as radiation oncology physics education and training.

\subsection{Potential applications of LLMs in radiation oncology physics}
Recent advancements in exploring potential applications of LLMs in radiation oncology physics have focused on auto-contouring, dose prediction and treatment planning. For auto-contouring, LLMs have been utilized to extract electronic medical records (EMR) text data and align them with the image embeddings of the mixture-of-experts model to improve the performance of the target volume contouring for radiation therapy~\cite{RAJENDRAN2025230}. In addition, LLMs have also been used to extract text-based features and incorporated them into vision transformer to help improve the target delineation results~\cite{oh2024mixturemulticenterexpertsmultimodal}. In dose prediction, LLM have been explored to encode knowledge from prescriptions and interactive instructions from clinicians into neural networks to enhace the prediction of dose-volume histograms (DVH) from medical images~\cite{https://doi.org/10.1002/mp.17416}. Regarding treatment planning, GPT-4V has been investigated for evaluating dose distribution and DVH and assisting with the optimization of the treatment planning~\cite{liu2025automatedradiotherapytreatmentplanning}. Furthermore, an LLM-based multi-agent system has also been developed to mimick the workflow of dosimetrists and medical physicists to generate text-based treatment plans~\cite{Wang_2025}. Collectively, these advancements highlight the transformative potential of LLMs in radiation oncology physics, offering potential improvements in efficiency and outcomes.

\subsection{Possible further improvement of LLMs in radiation oncology physics}
The performance of LLMs on radiation oncology physics, although encouraging, still requires further improvement due to two primary factors. First, radiation oncology physics represents a very specialized domain characterized by both the complexity of physics concepts and specific clinical contexts, neither of which was extensively represented in the general datasets used during the initial pre-training of these models. Second, existing LLMs still encounter difficulties with reasoning tasks specific to radiation oncology physics, indicating a need for enhanced general reasoning capabilities. To address these limitations, further studies could explore strategies of fine-tuning existing LLMs using specialized medical physics domain datasets with clinical contexts. Such fine-tuning would likely enable the models to better capture the complexities and contextual details of the domain, enhancing their accuracy and practical clinical utility in medical physics tasks. Additionally, to improve reasoning capabilities, techniques such as chain-of-thought, which encourages models to articulate intermediate reasoning steps explicitly, and reinforcement learning, which optimizes model responses in desired patterns, could be investigated~\cite{shao2024deepseekmathpushinglimitsmathematical}.

\subsection{Limitations}
Although the LLMs evaluated in this study exhibit expert-level performance on radiation oncology physics questions, such results do not directly translate to effectiveness in practical clinical tasks like treatment planning and delivery. This limitation arises from differences between theoretical examinations and practical clinical applications. Clinical scenarios encountered in radiation oncology are inherently more complex, context-dependent, and require integrating multiple sources of clinical and patient-specific data, whereas theoretical examinations often have clearly defined questions and objective answers. Consequently, strong performance in controlled question-answering tasks may not effectively transfer to real-world contexts, which frequently involve ambiguity, uncertainty, and nuanced clinical judgment. Additionally, clinical decision-making encompasses not only physics-based calculations but also multidisciplinary collaboration, patient safety considerations, regulatory compliance, and human factors in clinical workflows. Therefore, although the evaluated models demonstrate promise in foundational physics knowledge, caution must be exercised when inferring their direct clinical utility.

\section{Conclusion}
We evaluated recently released LLMs using a method that randomly shuffled the answer options of radiation oncology physics questions. Our results demonstrated that these models achieved expert-level performance on these questions, with some even surpassing human experts with a majority vote. However, when the correct answer options were replaced with "None of the above," all models exhibited a steep decline in performance, suggesting room for improvement. Employing the technique of explain-first and step-by-step instruction prompt enhanced the reasoning abilities of LLaMA 3.1 (405B), Gemini 1.5 Pro, and Claude 3.5 Sonnet. 

\section{Acknowledgments}
This research was supported by the National Cancer Institute (NCI) R01CA280134, the Eric \& Wendy Schmidt Fund for AI Research \& Innovation, The Fred C. and Katherine B. Anderson Foundation, and the Kemper Marley Foundation.



\printbibliography

\end{document}